# Artificial Skin Ridges Enhance Local Tactile Shape Discrimination

Saba Salehi, John-John Cabibihan[§], Shuzhi Sam Ge

Social Robotics Lab, Interactive and Digital Media Institute (IDMI) & Department of Electrical and Computer Engineering, National University of Singapore, 21 Heng Mui Keng Terrace, 119613 Singapore

[§]Corresponding author

Email addresses:

 SS: saba.salehi@gmail.com

 JJC: elecjj@nus.edu.sg

 SSG: samge@nus.edu.sg

# Abstract

**Background**

One of the fundamental requirements for an artificial hand to successfully grasp and manipulate an object is to be able to distinguish different objects' shapes and, more specifically, the objects' surface curvatures.

**Methods**

In this study, we investigate the possibility of enhancing the curvature detection of embedded tactile sensors by proposing a ridged fingertip structure, simulating human fingerprints. In addition, a curvature detection approach based on machine learning methods is proposed to provide the embedded sensors with the ability to discriminate the surface curvature of different objects. For this purpose, a set of experiments were carried out to collect tactile signals from a 2 × 2 tactile sensor array, then the signals were processed and used for learning algorithms. To achieve the best possible performance for our machine learning approach, three different learning algorithms of Naïve Bayes (NB), Artificial Neural Networks (ANN), and Support Vector Machines (SVM) were implemented and compared for various parameters.

**Results**

The most accurate method was selected to evaluate the proposed skin structure in recognition of three different curvatures. The results showed an accuracy rate of 97.5% in surface curvature discrimination.

# Introduction

The sense of touch is one of the fundamental sensations that give humans the ability to learn about objects' characteristics such as size, shape, surface texture, stiffness, and temperature [1-2]. Tactile sensing is also essential for detecting physical contacts [3] and it can effectively assist humans in object grasping and manipulation by providing information about the contact configuration [4]. Due to different shapes of objects, different forces and pressure patterns are generated. Therefore, the tactile information of the generated force patterns can be used to reproduce the knowledge of objects' shapes, which are required for grasping and manipulation in dexterous robotic hands. Specifically, it is the information of local shapes such as surface curvature of the objects that best determines the orientation and configuration of the fingers and the required grip force for a steady and reliable grasping [5]. Moreover, local tactile shape information can be used to determine the finger motion that will reach feasible grasping locations [6]. In the field of robotics, the perception of the tactile sensory information required for dexterous manipulation has fallen behind the mechanical capability of the existing hands [7-8], hence one of the primary requirements of robotics is to acquire an integrated knowledge of how tactile information is encoded and transmitted at various stages of interaction in order to design an effective robotic hand [8].

The current study is motivated by the lack of tactile information perception in artificial hands and the effective role of objects' local shape recognition in dexterous grasping and manipulation. It presents an approach for prosthetic/intelligent robotic hands to discriminate objects based on their surface shapes by employing machine learning methods. The main focus of the study is to propose a novel skin structure, which mimics the human fingerprints for artificial hands. The skin structure is ridged and aims to improve the efficiency of local shape detection for embedded tactile sensors. The next section reviews some of the most important studies on tactile shape recognition and fingerprint features.

## Background

**Tactile Shape Recognition: Global and Local Features**

Object recognition for robotic hands can be accomplished through a variety of object properties, such as geometric properties, including size and shape, or material properties, including texture, compliance, and temperature [9]. Object identification using tactile sensing has been shown to be accurate and quick, mostly in recognition through material properties [10]. However, emphasis of the researchers in this field has been mainly on analyzing the geometric characteristics due to their sufficiency for highly efficient recognition tasks [11]. Some objects used in previous studies, such as novel three dimensional clay objects [12] and objects with different geometric shapes cut out of stiff foam material [13], could not be distinguished based on their material properties. Only their geometric characteristics were useful and distinct in recognition. Among the geometric features, shape is a complex property which can be studied either globally or locally by decomposing its constituting elements such as curves and angles [14]. The recognition of global shape is out of scope of the current study. Global shape could be better identified by the aid of vision sensing together with tactile sensing [9], as vision is very fast and accurate in the recognition of objects through spatial properties. However, there are many situations in daily activities in which people interact with objects without using their vision, such as fastening the buttons of a dress, seeking an object in pockets, shifting gears, adjusting knobs while driving, and finding light switches in the dark [10]. This demonstrates the strength of tactile sensing for the objects' shape recognition, specifically the objects' local shape. The local shape of an object is described as the local features of that object's surface such as angles, edges, and curvature. Among these, curvature is invariant with respect to translation or rotation of the object and represents an important requisite to humans' perception of shapes. Humans are capable of learning and distinguishing slight differences in the curvature of various surfaces [15]. Hence, curvature is an influential factor which contributes to humans' reliability in precise recognition of shapes [15].

Tactile perception concerning the recognition of objects has been addressed in the literature [16-19]. Gaston and Lozano-Perez [18] examined the process of integration of information received from several tactile sensors for object identification. The purpose was to first grasp an object, chosen from a set of known objects, and then recognize the object, its position, and its orientation through a simple procedure with low computational cost. The information was obtained locally and was limited to the position of only a few contact points and a limited range of surface normal at the contact points. They were able to achieve the goal by using simple interpretation tree and pruning mechanisms.

Allen and Roberts [17] considered a database consisting of only six objects and presented a method for recognition of the objects using tactile image data. The gross shape of each object was first recovered through minimization of an error term. Next, the obtained parameters were utilized in the matching stage to fit the best object from the database to the parameters. Allen and Michelman [16], proposed an approach which recognized the objects' shapes through a set of exploratory procedures. The approach was to initially determine gross object shape and then to use a hypothesis-and-test method to generate more detailed information about an object. The sensory data that they used in their experiments were in the form of tactile images which were collected by active tactile sensing. Recently, Gorges et al. [19] presented an algorithm to classify the objects directly from the finger position and tactile patterns measured by an anthropomorphic robot hand. The purpose was to perceive the partial shape of objects by enfolding them successively using the hand. The sensory input data were tactile images and machine learning methods such as self organizing map and Bayes classifiers were used. They were able to recognize objects but not with a high accuracy. They further employed planar sensor surface which adapted its orientation passively to the object surface and gathered additional information for a better object recognition [19].

In addition to the studies of object recognition, others have also focused on active tactile sensing for local shape identification of objects, mostly on curvature detection. Okamura and Cutkosky [2] proposed an approach to define and identify surface curvature features using spherical robotic fingertips. They utilized various sensory data such as contact location, surface normal direction, and fingertip center position and concluded that not all of these information are needed for object shape recreation e.g., contact location.

An overview of the previous works on the tactile object discrimination indicates that these studies can be categorized into two major classes. These are works that have investigated the human tactile system and those that have implemented the knowledge on tactile systems to design and develop dexterous robotic hands. Our study belongs to the latter category. It addresses the neglected issue of tactile information perception for robotic/prosthetic hands in objects' local shape detection. Most of the earlier works were limited to a bounded number of known objects for recognition. Moreover, they have usually utilized the tactile images as their sensory input data. Higher memory usage for storing and analysis of images can decrease the speed of processing. In light of these, we propose an algorithm which is able to discriminate the local surface shape for unlimited number of objects using the force signals generated from tactile sensors.

**Fingerprint Features for Local Shape Discrimination**
Several studies have shown that fingerprints are effective in enhancing the recognition proficiency for both human and dexterous robotic hands. Fingerprints are the tiny ridges on humans' fingertips. Scientists have suggested various purposes for these papillary ridges such as serving as a mechanical amplification stage for stress transduction [20] and improvement of the consistency of the friction coefficient under moist conditions [21]. Vasarhelyi et al. [20] investigated the effect of fingerprints to assess their sensitivity to certain types of indentation and found out that besides enhancing the sensitivity, the modified geometries directly encode different properties of the indentation. Embedded tactile sensors can benefit from these results by increasing the sensor's sensitivity to different indentation types without reducing the level of protection.

It has been reported that fingerprints have the capability of reinforcing adhesion of the finger pads, which strengthens the object grasping task as well as acting as a magnifying layer [22]. According to Prevost et al. [23], fingerprints likely enhance the perception of texture by increasing vibrations, particularly the vibrations in the frequency range that best stimulates Pacinian corpuscles, in the skin as fingers rub across a textured surface. Later, they studied the role of fingerprints in the coding of tactile information and showed that fingerprints perform spectral selection and amplification of tactile information [22]. Therefore, it is advantageous to consider fingerprints in the form of small ridges for the skin of artificial hands following the results of previous studies. Majority of the earlier works have focused on embedded sensors with artificial ridges for sensitivity enhancement [24-25], vibration or slippage related grasping tasks [26]. However, our paper investigates the effect of the ridges on curvature discrimination.

The rest of the paper is organized as follows. In Section 3, the proposed skin structures and curvature discrimination approach for objects surface recognition are demonstrated. Section 4 provides the results for each step of the proposed approach and Section 5 concludes the paper.

## Curvature Discrimination Approach

Curvature is a parameter which is used to quantitatively express the curvature characteristic of objects. Curvature is defined as the amount by which an object is deviated from being flat [27]. In fact, curvature is represented by the changing direction of the unit tangent vector at each point of a curve. Consequently, the curvature of any of the segments of a straight line is zero as its unit tangent vector always maintains the same direction. One possible definition of curvature at each point on a curve is the reciprocal of the radius of the circle which passes through the point having the same conformance as the curve at that point [28].

To recognize the surface curvature of different objects, we developed a tactile interface. The proposed algorithm can be divided into three main phases, namely: (1) Indentation Experiments;

(2) Noise Elimination and Repeatability Assessment; and (3) Learning and Analysis. The first step is to conduct experiments with desired curvatures to collect the force signals that each curvature generates. These signals are later processed and utilized for curvature detection. The following subsection explains the first phase of our proposed procedure.

**Indentation Experiments**

We selected three main local surfaces of objects that are easily discriminated by humans. These were the flat, the edge, and the curved surfaces shown in Figure 1. These objects were made from the same material (steel) and were applied statically to a tactile sensor array to collect the force profiles that they generate. As the contact between the objects and the sensor is static, the stimuli need to be approximately the same size as the sensor. Otherwise, as the sensor is immobilized, large curved objects with large radii will be felt as flat. That is, only a small section of their surface, which has the characteristics of a flat surface, will be in contact with the small sensor. This situation introduces confusion between flat and curved objects in the stage of analysis. The tactile sensor array (TactoPad 2 × 2, TactoLogic Inc., Budapest, Hungary) that was used in the experiments was a 2 × 2 tactile sensor pad (Figure 2). The tactile sensor pad is a small contact-force mapping system, consisting of four (2 × 2) three-dimensional taxels in an array.

The sensing technology of the sensors is piezoresistive and the size of total active area is 4 mm × 4 mm with 5 mN/bit normal and 2 mN/bit shear sensitivity. This sensor can detect and measure the force values of the contact in both normal and tangential directions. It has been shown that the three-dimensional signals generated by this sensor provide the possibility for basic six Degree Of Freedom (DOF) spatial-temporal tactile measurements on arbitrary objects.

In order to investigate the effect of fingerprints, two skin covers with two different structures were considered [29-30]. One was flat [Figure 3(a)] and the other one was ridged [Figure 3(b)]. These skin covers, which were fabricated from silicone rubber (GLS 40, Prochima, Calcinelli, Italy), were designed to accommodate the rubber bumps from the TactoPad 2 × 2 sensor. The silicone material that was chosen was previously characterized and used for the synthetic skin of a cybernetic hand in [31-33]. This material type was also implemented for prosthetic finger phalanges with lifelike skin compliance [34].

The indenters were applied vertically on the skin covers and to the center of the four sensors on the Tactologic sensor pad to obtain their force profile. Figure 4 shows the cross-section of the indentation applied to the skin covers on the sensor pad.

The experimental pattern for all stimuli was to indent the sensor with the rate of 0.5 N/s (Phase A), hold the force for 2 s (Phase B), and then release with the same rate as the indentation (Phase C). The applied force at Phase B, which was approximately 2 N, is the maximum possible force that can be applied to TactoLogic sensor array such that no overload is imposed on the sensor. The three phases of A, B, and C are labeled in Figure 5.

Two separate experiments were carried out and the experimental data were used for learning and analysis. The data collected from the first experiment (Experiment 1, as shown in the left branch of Figure 6) was used for training and data collected from the second experiment (Experiment 2, as shown in the right branch of Figure 6) was held for testing, i.e., strictly no training or validation was done using this data.

In each of the experiments, two separate sets of data were collected for the two skin covers. In each set, forty indentations were made for each object. Hence, having three different objects and two skin covers, the total number of samples was 240 for each experiment.

**Noise Elimination and Repeatability Assessment**

The tactile sensor array has high sensitivity. Even the tiny movements of the stimuli during the indentation make the measured force signals noisy. Hence, for noise reduction, we need to choose an appropriate filter with proper characteristics, which are dependent to the noise frequency range of the collected data.

We designed a Finite Impulse Response (FIR) low pass filter to process our experimental data. For this purpose, we first determined the cut-off frequency. To achieve this, we formed the Fourier transform of the force signals and then obtained

Please cite this article in press as: Saba Salehi, John-John Cabibihan, Shuzhi Sam Ge, "Artificial Skin Ridges Enhance Local Tactile Shape Discrimination", *Sensors* 2011, 11(9), 8626-8642, doi 10.3390/s110908626- 11 - the frequency in which the power of the signal is reduced to half of the initial power. Then, based on the cut off frequency, which was about 1 Hz, the bandwidth of the pass-band and stop-band were drawn. It should be mentioned that a full window of collected data was used for processing in all phases.

Moreover, the repeatability of the measurements obtained through the proposed experimental procedure was assessed similar to the procedure in [35]. This was performed by calculating the pairwise Pearson cross-correlation of the measurement time series of all four sensors on the tactile sensor array. Pearson cross correlation coefficient is defined as follows:

$$r_{XY} = \frac{\sum_{i=1}^{n}(X_i - \bar{X})(Y_i - \bar{Y})}{\sqrt{\sum_{i=1}^{n}(X_i - \bar{X})^2}\sqrt{\sum_{i=1}^{n}(Y_i - \bar{Y})^2}} \tag{1}$$

where $\bar{X}$ and $\bar{Y}$ are the average values of the $X$ and $Y$ pair, respectively, and $n$ is the size of the measurement time series.

**Learning and Analysis**

The objective of the analysis is to develop a model that will be able to discriminate between flat, edge, and curved surfaces whenever an object comes into contact with the sensor. For this purpose, different machine learning algorithms were implemented to train a model based on the obtained data. The analysis for flat and ridged skin covers were carried out separately. Subsequently, the results of different methods were compared with each other, through model validation, in order to select the best one. The learning procedure, as shown in Figure 6, consists of feature extraction, training and validation, and testing. The details will be discussed in the following sections.

**Feature Extraction**

The purpose of the feature extraction is to select the optimum representations of a given dataset that contains the most useful information of the dataset and results in a desirable performance of a classifier. These features should be chosen carefully such that for each class they gather the most salient and distinguishable information. As our dataset contains force signals, the statistical features such as maximum value, mean, standard deviation, skewness, and kurtosis are the most informative features. Therefore, for each of the collected data, which comprises of signals from four sensors, these features were extracted and stored in a vector for classifier training. The features from all four sensors were considered as each sensor includes different information. Before applying the feature vector to a classifier, the data has to be normalized to a specific range such as [0,1] to prevent bias towards a special feature with a large value scale.

There are various normalization methods for data scaling such as Min-Max, Z-score, and decimal normalization. The algorithm used in this study is the Min-Max normalization [36], which is simple and fast, and is defined as follows:

$$x_{new} = \frac{x_{old} - x_{min}}{x_{max} - x_{min}} \left( \max_{desired} - \min_{desired} \right) + \min_{desired} \qquad (2)$$

where $\max_{desired}$ is the desired maximum value, i.e., 1, and $\min_{desired}$ shows the desired minimum value, i.e., 0.

**Training and Validation**

Once the features were extracted from the dataset and normalized, they were associated with a class label and then were utilized for model development. From the 40 samples corresponding to each stimulus with one of the skin covers, 30 were randomly chosen and used for model training (i.e., training dataset). The other 10 were held for validation (i.e., validation dataset). To perform the classifier training, we considered three machine learning approaches. These were Artificial Neural Networks (ANN) [37], Naïve Bayes (NB) [38], and Support Vector Machines (SVM) [39]. SVM is considered to be one of the algorithms, which achieve excellent performance. ANN is an earlier learning method with competitive performance compared to newer ones and NB is a simple method with poor average performance that occasionally performs exceptionally well [40]. These three algorithms belong to various categories of excellent, average, and poor performing methods, respectively. However, their performance can be significantly variable depending on different problems [40]. For neural network algorithm, Multi-Layer Perceptron (MLP) and for SVM, various Kernel functions with different corresponding parameters were investigated. These methods were applied to the training dataset to build a model and afterwards, the constructed models were evaluated using the validation dataset. To evaluate and compare the performance of different learning approaches, the accuracy rate was selected as the evaluation measurement metric defined as follows:

$$Accuracy\ rate = \frac{N_c}{N_t} \tag{3}$$

where *Nc* is the number of correctly predicted samples in validation phase and *Nt* is the total number of samples. Details of the accuracy rates for each skin cover and for different learning algorithms with different parameters are given in the Appendix.

**Testing**

In order to evaluate our approach, the test dataset was used to assess the recognition rate of objects with different curvatures. The features of this dataset were extracted following the same procedure explained in "Feature Extraction" Section and were fed to the selected classifier to obtain the results.

# Results

**Indentation Experiments**

Figure 7 shows the normal and tangential forces that were measured by each sensor on the TactoPad for flat and ridged covers when a flat object was indented on the skin surface.

Figure 8 presents three sets of raw experimental data for each of the three objects. The graphs are shown for both skin covers.

The experimental data for each skin cover with each of the objects is collected from different runs of the experiments carried out in the same condition. Each graph presents four responses corresponding to the resultant forces from the four sensors of the tactile sensor array. The four sensors on the sensor pad are three-dimensional taxels, which are embedded in an array. Due to different embedding errors, these four sensors measure different force values. However, it can be observed from the figure that the repeated measurements from each of the sensors have relatively same values for one skin cover with one surface. Moreover, the symmetrical characteristic of the responses shows that the indentation and release are made at the same rate.

**Noise Elimination and Repeatability Assessment**

Figure 9 displays a sample of filtered force signals for both flat and ridged skin covers indented with flat surface. Moreover, the average of Pearson cross-correlation coefficients is over 99%. This confirms that the collected measurements from the tactile sensor array with both skin covers are repeatable.

**Learning and Analysis**

Figure 10 presents the mean accuracy rate obtained for our validation dataset through various learning algorithms. The light gray bars represent the mean value of algorithms' accuracy rates with flat skin cover and the dark gray shows the mean accuracy rate for the ridged one. From the figure, it can be observed that the ridged skin cover performs better than the flat one in curvature recognition. This agrees with the previous results on capability of fingerprints in performing spectral selection and amplification of the tactile information about physical properties of the objects [22].

In addition, Figure 10 shows that SVM provides better performance compared to other two algorithms. The accuracy rates for different kernel functions, except for the polynomial kernel, are similar to each other for both flat and ridged covers. Hence, to select the best function, a trade-off between training and validation accuracies needs to be considered. Although SVM with linear kernel has the highest mean accuracy for ridged skin cover, the results of the Radial Basis Function (RBF) kernel is chosen as it has higher training accuracy rate (see Tables A1 and A2 for details). Moreover, RBF Kernel presents the highest accuracy rate for flat skin cover.

The performance of the learning algorithms, as well as skin structures, was further evaluated using a two-way analysis of variance (ANOVA) statistical test. A significance level of 0.05 was chosen. The results reveal that the learning algorithms (df = 4, F = 9.26, p = 0.0266), as well as skin structures (df = 1, F = 9.05, p = 0.0396) are significantly different. Moreover, the learning methods and skin structures perform independently as there was no interaction that was found between them. Separate t-tests were carried out to compare the performances of skin structures that resulted from each of the algorithms. Based on the tests' results, slight significant difference can be observed for ANN (p = 0.06) while no statistically significant differences can be found for NB (p = 0.5). However, SVM with linear (p < 0.001), Polynomial (p < 0.001), and RBF Kernel (p < 0.001) present significant differences between the ridged and flat skin covers. This indicates that in contrast to NN and NB, SVM can significantly discriminate the role of ridged cover compared to the flat cover. Additionally, based on the tests and Figure 10, it can be confirmed that ridged cover presents higher accuracy in surface curvature detection.

As the ridged skin cover and SVM algorithm with RBF kernel show superior performance over the other learning algorithms, for testing phase, only the test data corresponding to ridged skin cover was evaluated using SVM algorithm.

The surface curvatures of the test dataset were discriminated correctly with the accuracy rate of 97.5%. Table 1 indicates the confusion matrix, which shows the number of correctly-detected samples for each curvature. Furthermore, Figure 11 presents the comparison of the actual surface of objects and the predicted surfaces. Of the total of 120 samples collected with ridged skin cover for testing, the first 40 samples in the figure are curved objects, the second 40 are objects with an edge and the third 40 are flat objects.

## Conclusions

This paper aimed at proposing an approach to enhance local shape discrimination through tactile sensing. For this purpose, a novel ridged skin cover was introduced to simulate the effect of fingerprints for artificial hands and was compared with a flat skin cover. The results indicated that the performance of the proposed structure for the ridged skin cover is better than the flat cover in curvature discrimination. The small ridges, according to [22], can perform the spectral selection of force signals, which include information about local shape of objects.

We presented an offline surface curvature detection method based on machine learning algorithms. The proposed method included: (1) indentation experiments; (2) noise elimination and repeatability assessment; and (3) learning and analysis. The curvature detection method was evaluated for both ridged and flat skin structures. It was shown that it is possible to train a model using the tactile sensory data received from a tactile sensor, and then apply the learned model to new sensory data in order to recognize their local shape.

Moreover, it was observed that although there may not be any significant visual differences between the force signals for different curvatures, the selected statistical features can highlight the differences well. These features were used in recognition procedure for learning algorithms. Among various machine learning methods, support vector machine (SVM) showed to be the most efficient algorithm for classifying the force profiles of objects' surfaces. The results from the algorithm evaluation demonstrated that if the parameters of SVM are tuned accurately, a high accuracy rate can be expected in objects' surface discrimination. Finally, the average Pearson cross correlation coefficient value, which was over 99%, confirmed that the experiments and the measurements used for learning and analysis were repeatable.

## Acknowledgements

This work was supported by the Singapore Ministry of Education Academic Research Fund Tier1 Grant No. R-263-000-576-112.

## References

1. Cutkosky M, Howe R, Provancher W: **Force and tactile sensors.** In *Handbook of Robotics.* Berlin Heidelberg: Springer; 2008: 455-476
2. Okamura AM, Cutkosky MR: **Feature detection for haptic exploration with robotic fingers.** *The International Journal of Robotics Research* 2001, **20:**925.
3. Cheng; M-Y, Lin; C-L, Lai; Y-T, Yang; Y-J: **A Polymer-Based Capacitive Sensing Array for Normal and Shear Force Measurement.** *Sensors* 2010, **10:**10211-10225.
4. Tegin J, Wikander J: **Tactile sensing in intelligent robotic manipulation–a review.** *Industrial Robot: An International Journal* 2005, **32:**64-70.
5. Jenmalm P, Dahlstedt S, Johansson RS: **Visual and tactile information about object-curvature control fingertip forces and grasp kinematics in human dexterous manipulation.** *Journal of Neurophysiology* 2000, **84:**2984.
6. Fearing R: **Simplified grasping and manipulation with dextrous robot hands.** *Robotics and Automation, IEEE Journal of* 1986, **2:**188-195.
7. Carrozza M, Cappiello G, Micera S, Edin B, Beccai L, Cipriani C: **Design of a cybernetic hand for perception and action.** *Biological Cybernetics* 2006, **95:**629-644.
8. Dahiya RS, Metta G, Valle M, Sandini G: **Tactile sensing—from humans to humanoids.** *Robotics, IEEE Transactions on* 2010, **26:**1-20.
9. Hatwell Y, Streri A, Gentaz E: *Touching for knowing.* Amsterdam: John Benjamins Publishing Company; 2003.
10. Klatzky R, Lederman S: **The haptic identification of everyday life objects.** In *Touching for knowing: Cognitive psychology of haptic manual perception.* Edited by Hatwell Y, Streri A, Gentaz E. Amsterdam: John Benjamins Publishing Company; 2003: 105-122

11. James TW, Kim S, Fisher JS: **The neural basis of haptic object processing.** *Canadian Journal of Experimental Psychology* 2007, **61:**219-229.
12. James TW, Humphrey GK, Gati JS, Servos P, Menon RS, Goodale MA: **Haptic study of three-dimensional objects activates extrastriate visual areas.** *Neuropsychologia* 2002, **40:**1706-1714.
13. Zhang M, Weisser VD, Stilla R, Prather S, Sathian K: **Multisensory cortical processing of object shape and its relation to mental imagery.** *Cognitive, Affective, and Behavioral Neuroscience* 2004, **4:**251-259.
14. Gentaz E, Hatwell Y: **Haptic processing of spatial and material object properties.** In *Touching for knowing.* Edited by Hatwell Y, Streri A, Gentaz E. Amsterdam: John Benjamins Publishing Company; 2003: 123-159
15. Kappers A, Koenderink JJ, Lichtenegger I: **Haptic identification of curved surfaces.** *Perception and psychophysics* 1994, **56:**53.
16. Allen P, Michelman P: **Acquisition and interpretation of 3-D sensor data from touch.** *Robotics and Automation, IEEE Transactions on* 2002, **6:**397-404.
17. Allen PK, Roberts KS: **Haptic object recognition using a multi-fingered dextrous hand.** In *Book Haptic object recognition using a multi-fingered dextrous hand* (Editor ed.^eds.). pp. 342-347. City: IEEE International Conference on Robotics and Automation; 2002:342-347.
18. Gaston PC, Lozano-Perez T: **Tactile recognition and localization using object models: The case of polyhedra on a plane.** *Pattern Analysis and Machine Intelligence, IEEE Transactions on* 2009**:**257-266.
19. Gorges N, Navarro SE, Goger D, Worn H: **Haptic object recognition using passive joints and haptic key features.** In *Book Haptic object recognition using passive joints and haptic key features* (Editor ed.^eds.). pp. 2349-2355. City: IEEE International Conference on Robotics and Automation; 2010:2349-2355.
20. Vásárhelyi G, Ádám M, Vázsonyi É, Bársony I, Dücso C: **Effects of the elastic cover on tactile sensor arrays.** *Sensors and Actuators A: Physical* 2006, **132:**245-251.
21. Cutkosky M, Jourdain J, Wright P: **Skin materials for robotic fingers.** In *Book Skin materials for robotic fingers* (Editor ed.^eds.). pp. 1649-1654. City: IEEE International Conference on Robotics and Automation; 1987:1649-1654.
22. Scheibert J, Leurent S, Prevost A, Debregeas G: **The role of fingerprints in the coding of tactile information probed with a biomimetic sensor.** *Science* 2009, **323:**1503.
23. Prevost A, Scheibert J, Debrégeas G: **Effect of fingerprints orientation on skin vibrations during tactile exploration of textured surfaces.** *Communicative & Integrative Biology* 2009, **2:**422.
24. Gerling GJ: **SA-I mechanoreceptor position in fingertip skin may impact sensitivity to edge stimuli.** *Applied Bionics and Biomechanics* 2010, **7:**19-29.
25. Zhang Y: **Sensitivity enhancement of a micro-scale biomimetic tactile sensor with epidermal ridges.** *Journal of micromechanics and microengineering* 2010, **20:**085012.

26. Damian DD, Martinez H, Dermitzakis K, Hernandez-Arieta A, Pfeifer R: **Artificial Ridged Skin for Slippage Speed Detection in Prosthetic Hand Applications.** In *Book Artificial Ridged Skin for Slippage Speed Detection in Prosthetic Hand Applications* (Editor ed.^eds.). pp. 904-909. City; 2010:904-909.
27. Kline M: *Calculus: an intuitive and physical approach.* Dover Pubns; 1998.
28. Farin G, Sapidis N: **Curvature and the fairness of curves and surfaces.** *IEEE Computer Graphics and Applications* 1989**:**52-57.
29. Cabibihan J-J, Ge SS, Salehi S, Jegadeesan R, Hakkim HA: **Apparatuses, Systems, and Methods For Prosthetic Replacement Manufacturing, Temperature Regulations and Tactile Sense Duplication.** PCT application number PCT/SG2011/000255, Filing Date: 15 July 2011 (Pending).
30. Cabibihan J-J, Ge SS, Salehi S: **Apparatus, System, and Method for Tactile Sense Duplication For Prosthetic/Robotic Limbs with Ridged Skin Cover.** US Provisional Patent 61/370,640, 4 August 2011 (Converted).
31. Cabibihan J-J: **Patient-Specific Prosthetic Fingers by Remote Collaboration–A Case Study.** *PLoS ONE* 2011, **6:**e19508.
32. Cabibihan JJ, Pattofatto S, Jomaa M, Benallal A, Carrozza MC: **Towards humanlike social touch for sociable robotics and prosthetics: Comparisons on the compliance, conformance and hysteresis of synthetic and human fingertip skins.** *International Journal of Social Robotics* 2009, **1:**29-40.
33. Edin B, Ascari L, Beccai L, Roccella S, Cabibihan JJ, Carrozza M: **Bio-inspired sensorization of a biomechatronic robot hand for the grasp-and-lift task.** *Brain research bulletin* 2008, **75:**785-795.
34. Cabibihan JJ, Pradipta R, Ge SS: **Prosthetic finger phalanges with lifelike skin compliance for low-force social touching interactions.** *Journal of NeuroEngineering and Rehabilitation* 2011, **8**.
35. Oddo CM, Beccai L, Felder M, Giovacchini F, Carrozza MC: **Artificial roughness encoding with a bio-inspired MEMS-based tactile sensor array.** *Sensors* 2009, **9:**3161-3183.
36. Jain A, Nandakumar K, Ross A: **Score normalization in multimodal biometric systems.** *Pattern recognition* 2005, **38:**2270-2285.
37. Haykin S: *Neural networks: a comprehensive foundation.* Prentice Hall PTR Upper Saddle River, NJ, USA; 1994.
38. Friedman N, Geiger D, Goldszmidt M: **Bayesian network classifiers.** *Machine learning* 1997, **29:**131-163.
39. Cortes C, Vapnik V: **Support-vector networks.** *Machine learning* 1995, **20:**273-297.
40. Caruana R, Niculescu-Mizil A: **An empirical comparison of supervised learning algorithms.** In. ACM; 2006: 161-168.

# Figures

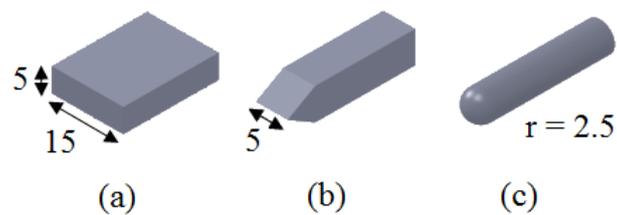

(a)      (b)      (c)

**Figure 1 -** (a) Experimental object with flat surface of 15 × 5 mm2. (b) Sharp edge in a line figure with the length of 5 mm and thickness of 0.1 mm. (c) Convex curved surface with a curvature of 0.4 mm−1 (radius of 2.5 mm).

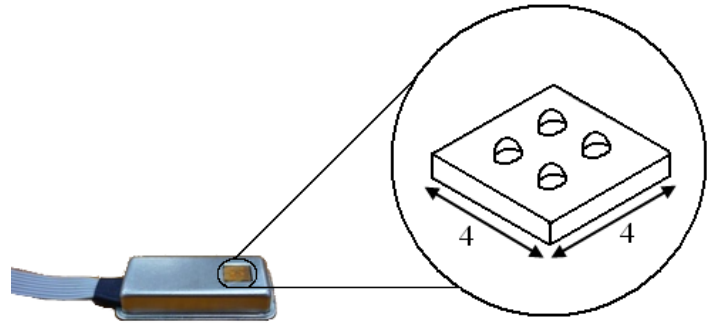

**Figure 2 -** Tactile sensor array. The sensor pad consists of four sensors on a 4 mm × 4 mm active area. Each sensor measures the contact force in normal and tangential directions.

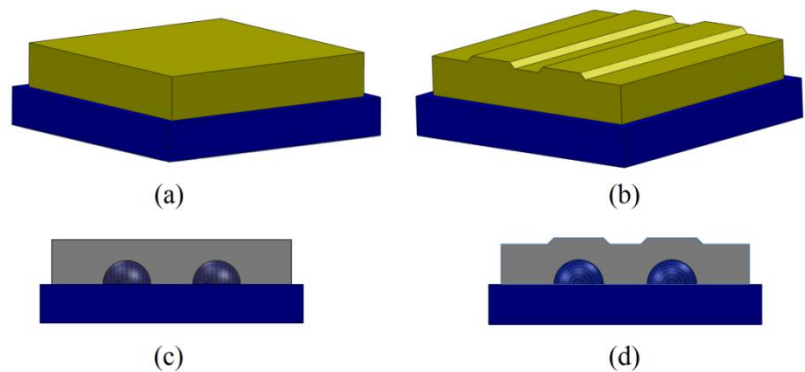

**Figure 3 -** (a) Perspective view of the flat skin cover that was included in the experiments as a baseline to verify the effect of ridges. (b) Perspective view of the ridged skin cover on the tactile sensor array. (c) Side view of the flat skin cover on tactile sensor array. (d) Side view of the ridged skin cover.

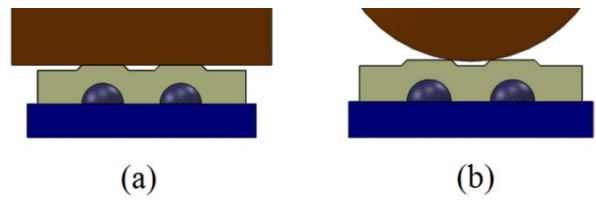

**Figure 4 -** Schematic of the indentation process. (a) Flat surface being applied to the ridged skin cover. (b) Curved surface being applied to the ridged skin cover.

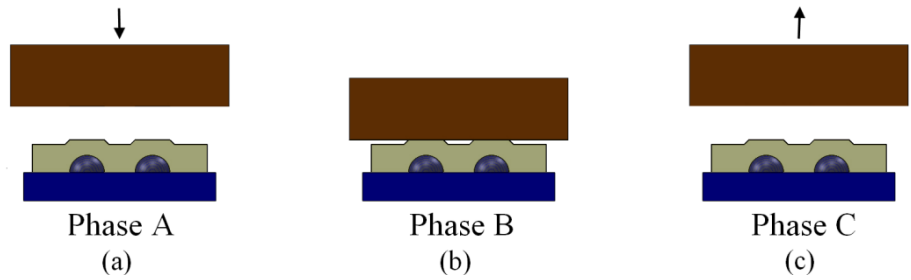

**Figure 5 -** Experimental pattern for indentation: (a) Phase A: Indent the sensor with the rate of 0.5 N/s. (b) Phase B: Hold the force for 2 s. (c) Phase C: Release with the same rate as the indentation.

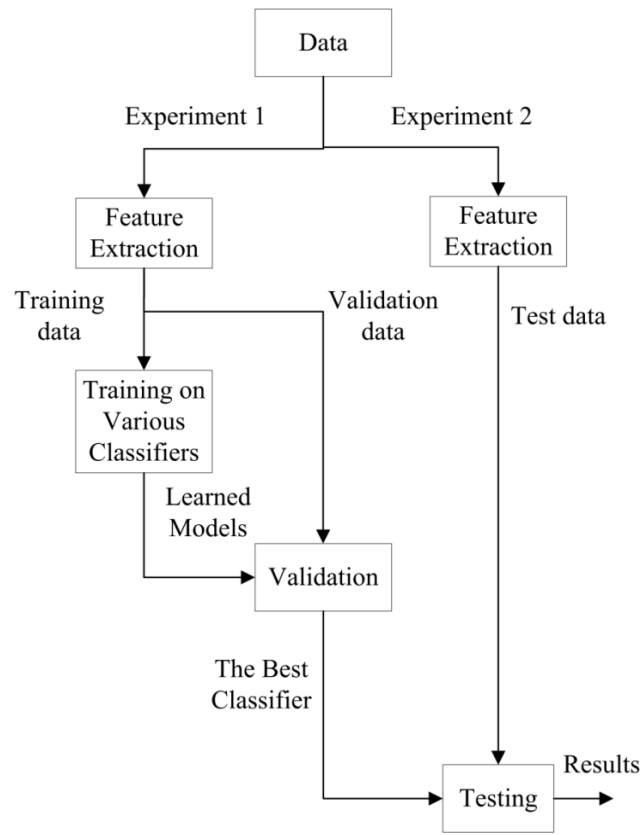

**Figure 6 -** Learning procedure. The left path shows classifier training using the collected data from experiment 1 and the right path displays testing of the trained model using collected data from experiment 2.

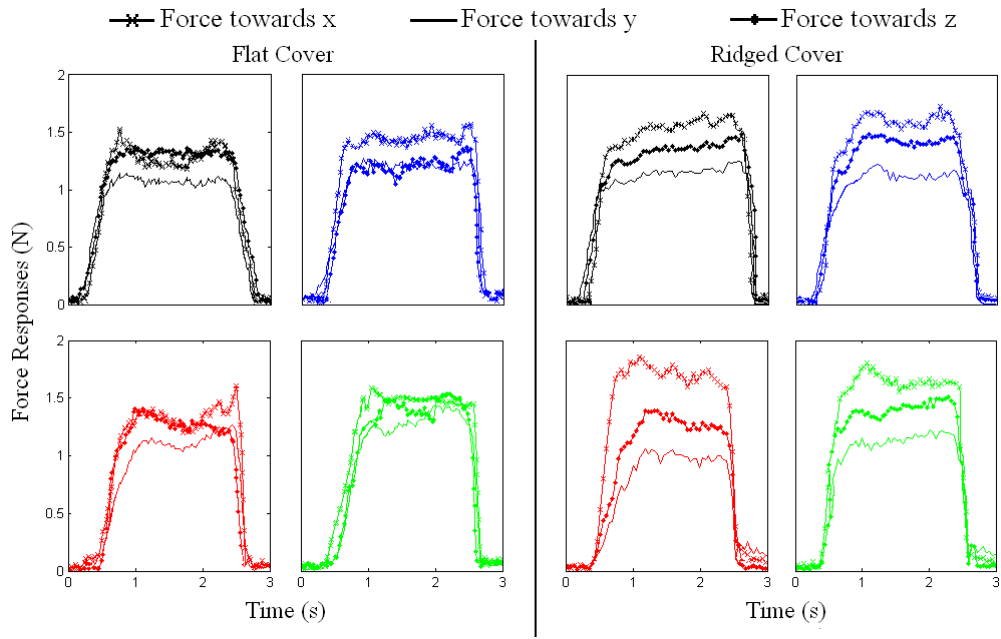

**Figure 7 -** Normal and tangential forces collected from all four sensors for the flat surface with flat and ridged covers.

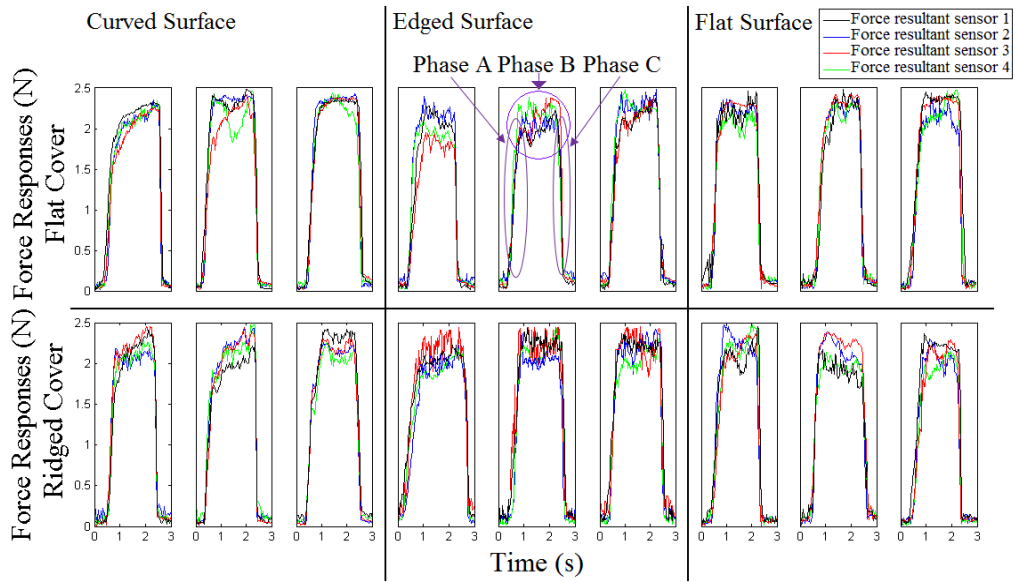

**Figure 8 -** Sample runs of the experiment for flat and ridged skin covers. The responses correspond to the resultant forces from the four sensors of the tactile sensor array.

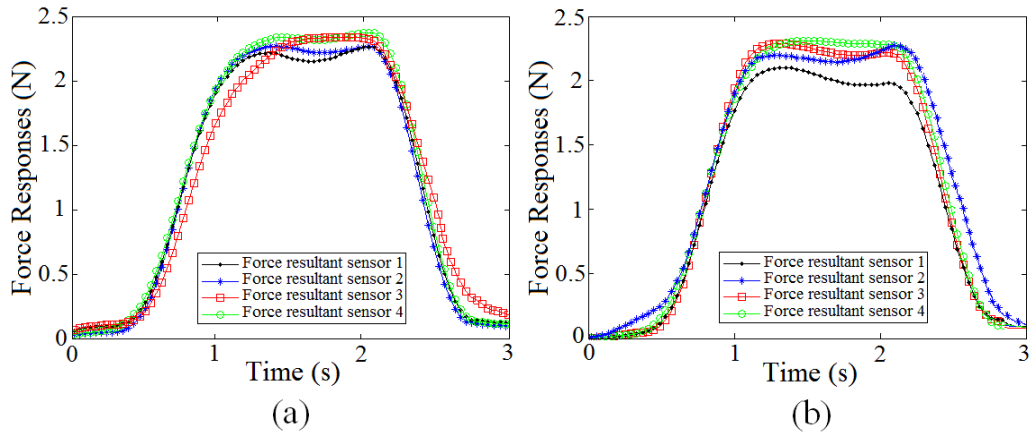

**Figure 9 -** (a) Force resultants collected from four sensors on the tactile sensor array with flat skin cover and flat surface, (b) Force resultants collected from four sensors on the tactile sensor array with ridged skin cover and flat surface.

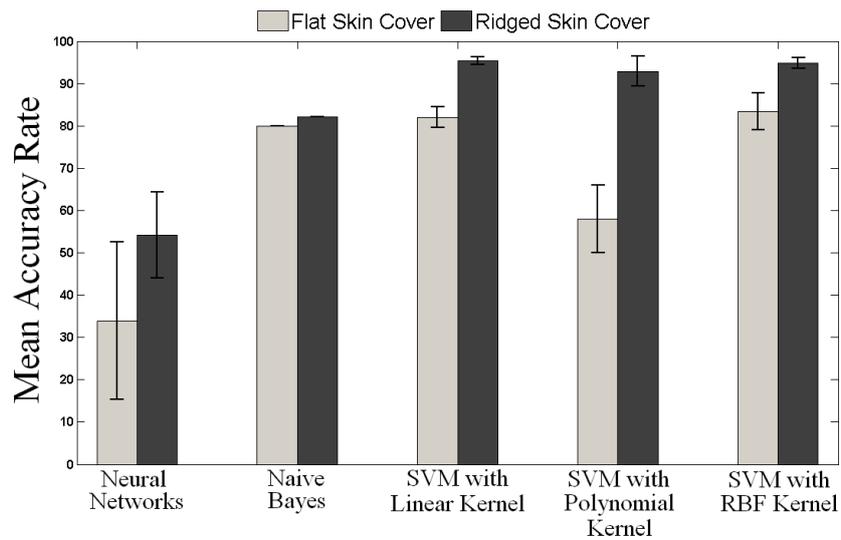

**Figure 10 -** Comparison of various machine learning algorithms. The light gray bars represent the mean value of algorithms' accuracy rates with flat skin cover and the dark gray shows the mean accuracy rate for the ridged one.

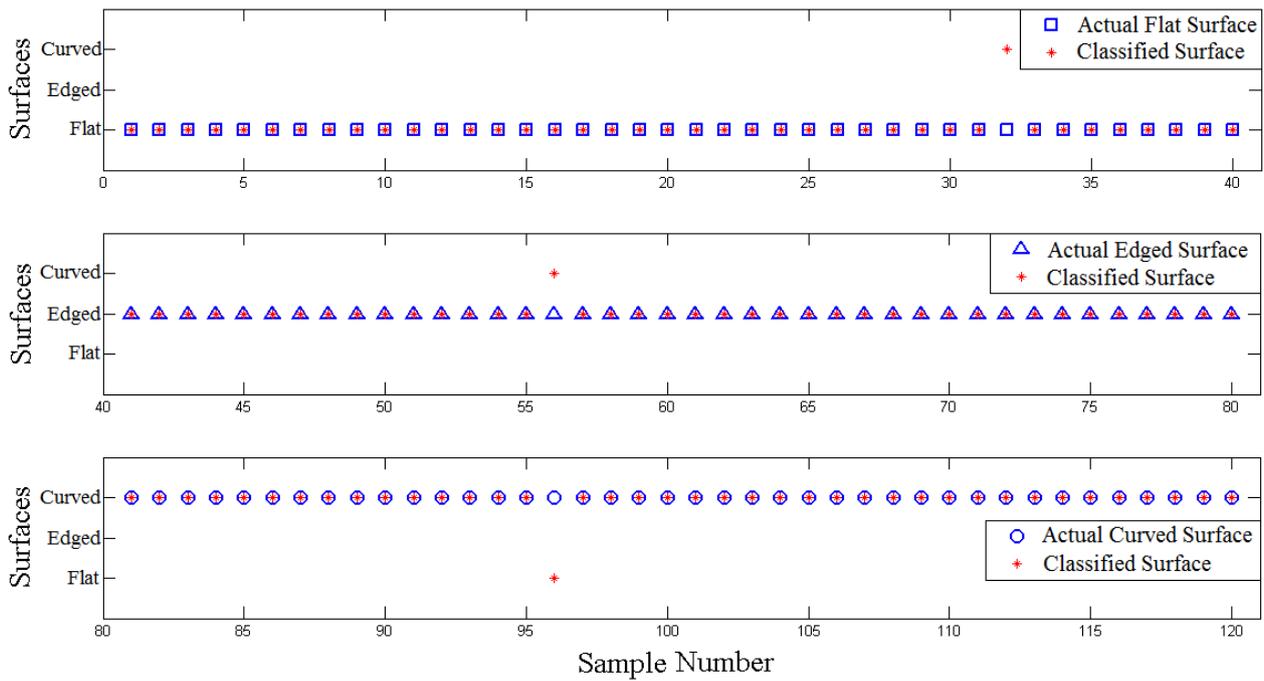

**Figure 11 -** Comparison of the actual surface of objects and the classified surface results. The top graph corresponds to the samples of the flat surface (samples 1–40), the middle graph shows the samples of the object with an edge (samples 41–80), and the bottom graph displays the samples of the curved surface (samples 81–120).

## Tables

**Table 1 -** Confusion Matrix

| | | Predicted Surfaces | | | |
|---|---|---|---|---|---|
| | | Curved | Edged | Flat | Total |
| Actual Surfaces | Curved | 39 | 0 | 1 | 40 |
| | Edged | 1 | 39 | 0 | 40 |
| | Flat | 1 | 0 | 39 | 40 |
| | | 41 | 39 | 40 | 120 |

# Appendix

**Table A1** – Accuracy rate results of different classification approaches for flat skin cover.

| Classifiers | Training Accuracy | | | | | Validation Accuracy | | | | |
|---|---|---|---|---|---|---|---|---|---|---|
| Naïve Bayes | 81.72% | | | | | 80% | | | | |
| Neural Network (MLP) | n[a] = 1 | n = 2 | | n = 3 | n = 5 | n = 1 | n = 2 | | n = 3 | n = 5 |
| | 33.3% | 30% | | 63.3% | 30% | 30% | 25.6% | | 53.3% | 26.7% |
| SVM with Linear Kernel | C[b] = 0.3 | C = 1 | C = 5 | C = 10 | C = 100 | C = 0.3 | C = 1 | C = 5 | C = 10 | C = 100 |
| | 88.3% | 90.8% | 93.3% | 94.2% | 97.5% | 78.33% | 80.83% | 84.2% | 81.7% | 85% |
| SVM with Polynomial Kernel | C = 0.3 | C = 1 | C = 5 | C = 10 | C = 100 | C = 0.3 | C = 1 | C = 5 | C = 10 | C = 100 |
| $p^c$ = 2 | 72.5% | 79.2% | 86.7% | 90% | 94.2% | 52.5% | 65% | 77.5% | 77.5% | 81.7% |
| p = 3 | 62.5% | 71.7% | 82.5% | 85% | 92.5% | 41.7% | 46.7% | 67.5% | 68.3% | 81.7% |
| p = 4 | 50% | 61.7% | 72.5% | 78.3% | 88.3% | 37.5% | 47.5% | 50% | 57.5% | 74.2% |
| p = 5 | 43.3% | 53.3% | 69.2% | 71.7% | 85.8% | 34.2% | 42.5% | 46.7% | 46.7% | 61.7% |
| SVM with RBF Kernel | C = 0.3 | C = 1 | C = 5 | C = 10 | C = 100 | C = 0.3 | C = 1 | C = 5 | C = 10 | C = 100 |
| $\gamma^d$ = 0.1 | 81.7% | 86.7% | 91.7% | 92.5% | 98.3% | 78.3% | 75% | 84.2% | 83.3% | 88.3% |
| $\gamma$ = 0.3 | 85.8% | 92.5% | 94.2% | 95.8% | 98.3% | 77.5% | 80.8% | 85% | 85.8% | 90% |
| $\gamma$ = 0.5 | 90% | 92.5% | 95.8% | 97.5% | 99.2% | 79.2% | 84.2% | 83.3% | 86.7% | 90% |

[a] n shows the number of neurons in the hidden layer of a Multi-Layer Perceptron
[b] C represents the upper bound parameter of the SVM soft margin
[c] p is the degree of the Polynomial kernel
[d] $\gamma$ shows the parameter of SVM Radial Basis Function kernel

**Table A2 –** Accuracy rate results of different classification approaches for ridged skin cover.

| Classifiers | Training Accuracy | | | | | Validation Accuracy | | | |
|---|---|---|---|---|---|---|---|---|---|
| Naïve Bayes | 86.67% | | | | | 82.22% | | | |
| Neural Network (MLP) | $n^a$ = 1 | n = 2 | | n = 3 | n = 5 | n = 1 | n = 2 | n = 3 | n = 5 |
|  | 52.2% | 66.7% | | 61.1% | 56.7% | 40% | 66.7% | 60% | 50% |
| SVM with Linear Kernel | $C^b$ = 0.3 | C = 1 | C = 5 | C = 10 | C = 100 | C = 0.3 | C = 1 | C = 5 | C = 10 | C = 100 |
|  | 88.3% | 91.7% | 95.8% | 96.7% | 98.3% | 94.2% | 95% | 95.8% | 95% | 96.7% |
| SVM with Polynomial Kernel | C = 0.3 | C = 1 | C = 5 | C = 10 | C = 100 | C = 0.3 | C = 1 | C = 5 | C = 10 | C = 100 |
| $p^c$ = 2 | 81.7% | 81.7% | 85% | 87.5% | 94.2% | 94.2% | 94.2% | 95.8% | 95.8% | 96.5% |
| p = 3 | 71.7% | 71.7% | 71.7% | 78.3% | 86.7% | 93.3% | 92.5% | 95.8% | 95% | 96.5% |
| p = 4 | 65% | 65% | 65% | 65% | 83.3% | 90% | 90.8% | 93.3% | 93.3% | 96.5% |
| p = 5 | 59.2% | 59.2% | 59.2% | 59.2% | 65.8% | 84.2% | 85% | 89.2% | 91.2% | 95% |
| SVM with RBF Kernel | C = 0.3 | C = 1 | C = 5 | C = 10 | C = 100 | C = 0.3 | C = 1 | C = 5 | C = 10 | C = 100 |
| $\gamma^d$ = 0.1 | 86.7% | 88.3% | 92.5% | 95% | 99.2% | 91.7% | 93.3% | 95% | 95.8% | 96.5% |
| $\gamma$ = 0.3 | 87.5% | 89.2% | 95.8% | 96.7% | 100% | 93.3% | 95% | 95.8% | 95.8% | 95% |
| $\gamma$ = 0.5 | 88.3% | 91.7% | 96.7% | 98.3% | 100% | 95% | 95% | 95.8% | 95.8% | 93.3% |

[a] n shows the number of neurons in the hidden layer of a Multi-Layer Perceptron

[b] C represents the upper bound parameter of the SVM soft margin

[c] p is the degree of the Polynomial kernel

[d] $\gamma$ shows the parameter of SVM Radial Basis Function kernel